\documentclass{sig-alternate-10pt}

\usepackage{fancyhdr}
\usepackage{datetime}
\usepackage{amsmath}
\usepackage{graphicx}
\usepackage{textcomp}
\usepackage{epsfig}
\usepackage{multirow}
\usepackage{url}
\usepackage{color}
\usepackage{times}
\usepackage{balance}
\usepackage{rotating}
\usepackage{footnote}
\usepackage{listings}
\usepackage{paralist}

\usepackage[kerning=true]{microtype}

\begin{document}

\title{Towards a Stateful Forwarding Abstraction to Implement Scalable Network Functions in Software and Hardware}

\author{
\rm{ 
	Luca Petrucci$^{\ddagger}$,
	Nicola Bonelli$^{\ast}$,
	Marco Bonola$^{\ddagger}$,
	Gregorio Procissi$^{\ast}$,
	Carmelo Cascone$^{+}$,}\\ 
\rm{ 
	Davide Sanvito$^{+}$,
	Salvatore Pontarelli$^{\ddagger}$,
	Giuseppe Bianchi$^{\ddagger}$},
	Roberto Bifulco$^{\dagger}$\\
\rm{$^{\ast}$ CNIT/University of Pisa}
\rm{$^{\ddagger}$ CNIT/University of Rome Tor Vergata}\\
\rm{$^{+}$ CNIT/Politecnico di Milano}		
\rm{$^{\dagger}$ NEC Laboratories Europe}  
}

\maketitle

\begin{abstract}
An effective packet processing abstraction that leverages software or hardware acceleration techniques can simplify the implementation of high-performance virtual network functions. In this paper, we explore the suitability of SDN switches' stateful forwarding abstractions to model accelerated functions in both software and hardware accelerators, such as optimized software switches and FPGA-based NICs. In particular, we select an Extended Finite State Machine abstraction and demonstrate its suitability by implementing the Linux's iptables interface. By doing so, we provide the acceleration of functions such as stateful firewalls, load balancers and dynamic NATs. We find that supporting a flow-level programming consistency model is an important feature of a programming abstraction in this context. Furthermore, we demonstrate that such a model simplifies the scaling of the system when implemented in software, enabling efficient multi-core processing without harming state consistency.
\end{abstract}

\section{Introduction}
\label{sec:intro}

Network functions, commonly implemented using hardware middleboxes, are being transformed in virtualized software appliances that run on commodity servers~\cite{clickOS}. Network operators are supporting this trend~\cite{cord16}, usually called \emph{Network function Virtualization} (NFV)~\cite{nfv}. Virtual network functions (VNFs) have a number of advantages when compared to legacy hardware ones. They can be dynamically created, updated, migrated and run on commodity servers. However, developing VNFs for carriers is a hard task~\cite{NetBricks}. A relevant challenge is the need to meet strict requirements in terms of performance and reliability, while running in general purpose, multi-tenant (i.e., virtualized) environments. Furthermore, the required packet forwarding performance is a continuously rising bar. A commodity server is usually equipped with a couple of 10Gbps network interfaces, and 40Gbps interfaces are becoming common. Unfortunately, current general purpose systems' speed is not growing as fast as the network interfaces speed~\cite{moore2015}. Therefore, carriers increasingly consider the introduction of network accelerators, based on smart NICs, in their VNF designs~\cite{openNFP, agilio, netsume}. In fact, smart NICs help in supporting the implementation of hardware accelerators as required by future network functions\footnote{Such a need has been recently recognized also by standard organizations such as ETSI ISG NFV, which is defining hardware accelerators' capabilities in their specifications for NFV platforms~\cite{etsinfvifa002}.}, without waiting for the long implementation cycles experienced with traditional NICs~\cite{lavanyaNSDI15}.

An effective way to simplify the implementation of a high-performance network function is to separate the function's fast path from its control path~\cite{keshavRouters}. Here, the data and control planes separation of the OpenFlow's architecture is a relevant example~\cite{openflow}. The OpenFlow controller implements the slow path using high-level languages such as C, Java, Python, etc. The switch implements the fast path using a model that comprises a pipeline of match-action tables (MATs). Therefore, a network function's developer can rely on the already optimized implementation of the MAT abstraction, e.g., provided by a fast software switch~\cite{ovs-paper}, for implementing a high-performance function's fast path.

Unfortunately, OpenFlow (and similar SDN abstractions) is too constrained. In essence, only packet classification based on L2-L4 protocol headers and some header rewriting capabilities are exposed, limiting its applicability only to the definition of trivial VNFs~\cite{softflow}. 

\vspace{0.05in}
\noindent\textbf{Fast Path Abstractions}
Ideally, we would like to define a fast path abstraction that simplifies the implementation of high-performance VNFs, while enabling a seamless adoption of future hardware accelerators. In other words, we want to answer the following questions: is it possible to define a common abstraction to efficiently model software and hardware VNFs fast path functions? If such an abstraction exists, can it model a relevant set of functions? Can the abstraction support the performance, runtime reconfiguration and multi-tenancy requirements of NFV environments? 

An important observation is that the already mentioned OpenFlow's abstraction, based on MATs, is a good match for many of the requirements we need to fulfill. First, it can be implemented in software with high-performance~\cite{ovs-paper} and is supported by hardware implementations. As a matter of fact, NIC vendors are already leveraging OpenFlow-like models for the programming of NIC's functions~\cite{bcm57300}. Second, OpenFlow devices are configured at runtime by changing the flow entries in the MATs, without interrupting the device functions. This enables scenarios in which VNFs are dynamically deployed or migrated. Third, MAT-based abstractions have been proved effective to implement virtualization and multi-tenancy~\cite{flowvisor}.

Unfortunately, many of the abstractions that extend a MAT-based model to support more flexible functions target only hardware implementations (cfr. Sec.~\ref{sec:abstractions}). This is particularly true when the abstraction supports the definition of network functions that explicitly manage some sort of algorithmic state. That is, so-called stateful abstractions. As a result, there is either a lack of a suitable consistency model to describe stateful functions, or there is a consistency model that cannot be easily implemented (and scaled) in, e.g., a software environment based on general purpose CPUs. 

\vspace{0.05in}
\noindent\textbf{Contribution}
In this paper, we build on top of an extension of OpenState~\cite{openstate}, i.e., the Open Packet Processor (OPP)~\cite{bebaD23, bebaD24}, which maintains an OpenFlow-like model for programming network devices, but tweaks it for the implementation of an Extended Finate State Machine abstraction (EFSMs)~\cite{efsm}. That is, we believe that OPP may improve on the lack of flexibility of OpenFlow, while still retaining most of its previously outlined benefits. Furthermore, OPP introduces a flow-level state consistency model, which eases a scalable  implementation of the abstraction in software.

Therefore, we provide two main contributions. First, we demonstrate that the OPP abstraction is flexible and can in fact model a relevant set of commonly used network functions. In particular, we implement the Linux's iptables interface (cfr. Sec~\ref{sec:usecases}) on top of OPP. Supporting iptables has the following benefits: it enables the entire set of functions supported by iptables, such as stateful firewalls,  load balancers and dynamic NATs, to name a few; it provides an interface that is already commonly used~\cite{yoda} and which is well-known by developers, while still enabling the implementation of additional, perhaps more innovative, new functions\footnote{We remark that such an approach has been already proven successful in the past, since familiar abstractions enable one to re-use current implementations, policies and configurations without modification~\cite{nvp}.}. Second, we provide an OPP software implementation that can dynamically scale to run on multiple CPU cores, maintaining the consistency of the implemented stateful network functions (cfr. Sec~\ref{sec:implementation}). We identify in the OPP's flow level state consistency the key abstraction to enable a scalable software implementation. Since there is already an OPP implementation for the NetFPGA SUME~\cite{netsume}, our software implementation demonstrates that OPP can be implemented efficiently in both hardware and software. 

\section{Stateful Abstractions}
\label{sec:abstractions}

In this section, we briefly recap our requirements for a stateful forwarding abstraction for the data plane, and review currently available ones that suite our needs. Finally, we provide a primer of OPP, which is our selected abstraction.

\vspace{0.1in}
\noindent\textbf{Requirements}. 
As previously mentioned, we are interested in selecting an abstraction that is flexible enough to model several network functions' data plane while fulfilling the following requirements: (i) it is efficiently implementable in both software and hardware (i.e., smart NICs); (ii) it enables the deployment of new functions at runtime; (iii) it can support multiple concurrent functions. The OpenFlow's MAT-based abstraction fulfills the above three requirements, but fails to model an important set of network functions that require to keep algorithmic state. We refer to these set of functions as stateful functions. Therefore, a stateful abstraction is one that can model stateful functions.

Given the good match of a MAT-based abstraction with our requirements, it is reasonable to explore extensions of such abstraction. Hopefully, one could retain the required properties while adding the flexibility that is lacking in the current OpenFlow abstraction.

\vspace{0.1in}
\noindent\textbf{P4}. Recently, a number of proposals started to investigate stateful forwarding abstractions in the fast path, using a pipeline of MATs as base model. Most notably, the P4 language~\cite{p4} allows a programmer to describe stateful functions using a pipeline of configurable MATs, which is defined in a hardware independent manner. P4 can be then compiled to a number of \textit{targets}, for instance a smart NIC~\cite{flexNIC} or a software switch~\cite{pisces}. To describe stateful functions, P4 provides \textit{registers} that can be accessed within the MATs pipeline to keep state (and manipulate it). 

While the hardware independent nature of P4 makes it appealing, we found that none of the currently available versions of P4 \cite{P4spec10,P4spec11} defines a concurrency model that helps in managing the stored state. 
In other words, P4 cannot model assumptions on how the data (registers) is read or written and, therefore, it is hard to provide guarantees in terms of state consistency in actual implementations. This is a fundamental need when dealing with state that can be shared and manipulated concurrently, e.g., by different stages of a packet processing pipeline. In fact, such information is used to arbitrate and optimize memory accesses in actual implementations. For instance, NetASM~\cite{netasm}, an abstract intermediate representation for programmable pipelined data planes, captures this information by distinguishing between per packet and persistent states. 

As a result, in P4 the only way in which state consistency, hence algorithm correctness, is guaranteed is to assume that the pipeline processes one packet at a time. Such assumption is of course not acceptable, since pipelining of packets is required to provide high performance in hardware~\footnote{We are aware that the P4 Language Design Working Group is currently working on a concurrency model for the next release of the language. However, at the time of writing, this aspect is still work in progress and, to the best of our knowledge, we are not aware of any target hardware platform implementing such specification.}. 

\vspace{0.1in}
\noindent\textbf{POF}. POF~\cite{pof} enables the definition of flexible MATs, with actions that could be used to read and modify state. Since similar considerations to those of P4 apply here, in the interest of space, we do not describe in details POF in this paper.

\vspace{0.1in}
\noindent\textbf{Banzai}. Another relevant proposal is Domino/Banzai~\cite{domino}. Here, the authors propose both a Domain Specific Language (DSL), Domino, and a machine model, Banzai, for the definition and implementation at line rate of stateful packet processing functions. The Banzai machine exposes a set of specialized stateful processing instructions called ``atoms'', used to implement the action part of a MAT. Atoms have two main limitations: they must execute their operations atomically and cannot share state with other atoms. Differently from P4, a program written in Domino, if accepted by the Domino's compiler, offers strong consistency guarantees. The compiler is responsible to find a mapping of the portions of a Domino program to the available atoms. Depending on the expressiveness of the atoms, \cite{domino} shows that a fair amount of stateful functions can be expressed.

Unfortunately, the Banzai model has a few issues when applied to our context. First, the model is thought to work with line rate switches. That is, it is based on the assumption of having a sequential pipelined processor. In software, this assumption is not true anymore. 
As a result, unless only one of the CPU's cores is used to process sequentially all the packets, limiting scalability, the assumption that atom's state can be atomically updated would require a software implementation to resort to expensive resource locking techniques. That is, it would need to provide mutually exclusive access from different CPU's cores to the atom's state. Furthermore, such locking should happen for each processed packet and for each pipeline's stage.  Second, having a consistency guarantee based on the atomic execution of state updates frees the programmer from stating its assumptions on how state is accessed. This simplifies the programming of functions in Domino, but brings in the same issues mentioned for P4. That is, it leaves unspecified which portion of the data a given function reads or writes. Finally, at the time of writing, we are not aware of any Banzai machine implementation for smart NICs.

\subsection{Open Packet Processor}
The abstractions mentioned so far are all lacking explicit information about which part of the overall state is used for the implementation of a given algorithm. Nonetheless, in most of the cases each packet actually belongs to one of several different \textit{state contexts}. For instance, in the case of a stateful firewall, each bi-directional flow has its own state, e.g., new or established, and two flows don't usually affect each other state. As in the case of NetASM, using an abstraction that can identify and isolate these contexts helps in solving state consistency problems on a per-context level rather then on a global level.  

Notable examples of stateful forwarding abstractions that provide the identification of state contexts are FAST~\cite{fast} and OpenState~\cite{openstate}. Since the approaches are quite similar, we will be focusing on the latter, with which we are more familiar. Also, OpenState recently evolved into the Open Packet Processor (OPP), which provides a more powerful abstraction while maintaining the core of the OpenState ideas.

In OPP, Extended Finite State Machines (EFSM) are used to model stateful forwarding algorithms. We leave aside the details of the EFSM implementation, to which we will come back shortly, to point instead to an important property of the OPP's abstraction. That is, the need to explicitly define which part of the overall device's state a packet is going to modify during its processing. As in FAST, such operation is done by defining the state on a per-flow basis. An arbitrary combination of packet's header values\footnote{Including metadata such as the port the packet was received on.} is used to define the portion of the state such packet is going to (and allowed to) use. The state identified in this way is called \emph{flow context}. Furthermore and differently from FAST, in OPP a packet can read one flow context and update a different one. In fact, it is possible to specify  different combinations of packet's header values for the read and update operations of a flow context. This last property is named \emph{cross-flow state update}.

The explicit handling of flow contexts and the support for cross-flow state updates helps in the implementation of a number of functions. For instance, one could implement a simple L2 learning switch, or more complex applications, as we will show later presenting an implementation of the iptables' interface (Sec.~\ref{sec:usecases}). The rest of this section presents the OPP's machine model.

\subsubsection{Machine model}
\begin{figure*}[t]
	\centering
	\includegraphics[width=0.95\textwidth]{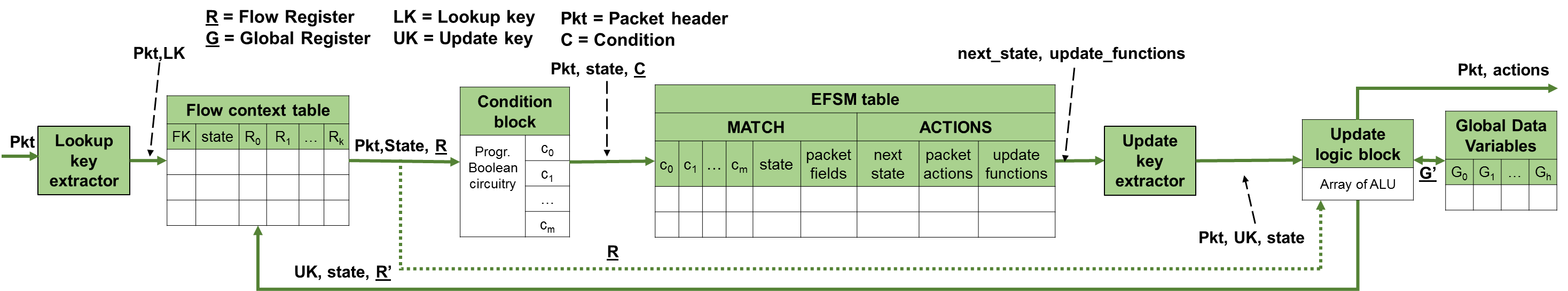}
	\vspace{-0.1in}
	\caption{Architecture of an OPP processing block}
	\label{fig:hl_arch}
	\vspace{-0.1in}
\end{figure*}

The OPP machine model extends the MATs pipeline model assumed by OpenFlow. MATs are substituted with \emph{stages}, which can be either stateless or stateful. A stateless stage is in fact an OpenFlow-like MAT. The pipeline processes packets' headers to define corresponding forwarding behaviors. OPP assumes packets headers are already parsed when passed to the pipeline, therefore, OPP can potentially leverage related work on programmable packet parsing and reconfigurable match tables~\cite{programmableParser, rmt}. A stateful stage (Fig.~\ref{fig:hl_arch}) adds a number of elements to a plain MAT.

 \noindent\textbf{Flow context}. First, when a packet header enters the stage, a \emph{lookup key extractor} builds a key that uniquely identifies a flow context for such packet. The extractor is programmed at runtime by specifying a list of relevant header fields and packet's metadata (e.g., the TCP/UDP 4-tuple or the \textit{in\_port}). The key is then used to extract the flow context from the \emph{flow context table}. The context includes a state label $s$ and an array of registers $\vec{R} = \{r_0, r_1, ..., r_{(k-1)}\}$. If no context is found for a given key, a default context is used (i.e., with all values set to 0). Furthermore, each flow context can be associated with a (hard or idle) timeout.

 \noindent\textbf{Conditions}. Second, the packet's header and metadata, which now include the flow context, are passed to a \emph{condition block}. The condition block can be programmed at runtime for the evaluation of up to $m$ expressions with mathematical operators like $>$, $<$, $=$. For example, conditions can be used to compare if a port number is greater than the value stored in a given flow context's register. 
The output of the condition block is a Boolean vector $\vec{C} = \{c_0, c_1, ..., c_{(m-1)}\}$, where $c_i=1$ if the $i$-th condition is true, otherwise $c_i=0$. 

\noindent\textbf{EFSM table}. Third, the packet's header and metadata, plus $\vec{C}$ and state label $s$, are passed to the \emph{EFSM table}. Such table is a typical MAT that supports ternary matching on the just mentioned values. For each entry in the table, a programmer can specify (i) a list of OpenFlow-like actions to be executed on the packet, (ii) the next state label $s$ in which the flow context shall be set, and (iii) a list of instructions to update the flow context registers $\vec{R}$. Furthermore, the update functions can also operate on the global variables $\vec{G} = \{G_0, G_1, ..., G_{(h-1)}\}$. Since the variables $\vec{G}$ are global, their read and update operations happen atomically. In effect, the EFSM table describes the transitions of the state machine.

\noindent\textbf{Context update}. Finally, the packet's header and metadata, the action to be applied, the update instructions and the new value of the state label are passed to the \emph{update logic block}. Here, an array of Arithmetic and Logic Units (ALUs) performs the required update instructions to update the values stored in both the flow context registers $\vec{R}$ and global registers $\vec{G}$, using arithmetic functions. Such functions can range from simple integer sums, for instance to update the value of a register representing a packet or byte counter, to more complex ones, e.g., floating point processing, depending on the specific implementation and required performance. The result of this block is then used to update the flow context identified by the key generated by the \emph{update key extractor}. Such extractor works in the same way of the lookup key extractor.

\subsubsection{Remarks}
The OPP model defines three different types of states, with corresponding consistency models.

First, the \emph{packet state}, which has a non-concurrent access model. That is, it can be only accessed and updated during the current packet processing. By definition, such state is created when a packet enters the stage and destroyed once the processing is complete. The packet state is practically described by the packet's metadata. Here, it is worth to notice that all the discussed abstractions, including OpenFlow, define the packet state. 

Second, the \emph{global state}, which is common to all the packets. This state exists for the entire life of the stage and is stored by the $\vec{G}$ registers. As in Banzai, in OPP the global state can be only updated with atomic operations. In practice, a programmer that uses global states should be well aware that the strong consistency guarantees may introduce performance overheads in some implementations. 

Third, OPP defines a \emph{flow state}. The flow state only exists for the life of a flow and is captured in the OPP's flow context. Here, the consistency is guaranteed on a per-flow basis, since the model does not allow concurrent accesses to the flow state. However, multiple packets accessing and modifying different flow states can be processed in parallel. 
Which packets can execute in parallel can be easily derived. Recall that the programmer has to explicitly specify beforehand the need to read and write a flow state during the processing of a packet. Such operation is concretely performed configuring the lookup and update key extractors.

Finally, notice that these states are all contained within one stage. In fact, only n bits of the packet metadata (i.e., the packet state) can be used as a mean to move state from one stage to the next.

\subsection{NetFPGA prototype}
\label{sec:proto}

An OPP hardware prototype is available for the NetFPGA SUME \cite{netsume}. Describing the details of the prototype is out of scope for this paper. However, we report a few important data that help putting the OPP abstraction in a more concrete perspective in terms of hardware requirements and constraints. First, the prototype can forward the packets of 4 10GbE ports at line rate. Providing a 156.25Mpps (Million packets per second) throughput. The system may support flow context consistency by adding queues before a stage. Such queues have to hold packets that can potentially access the same flow contexts concurrently. However, notice that: (i) the pipeline can be meanwhile used by other packets belonging to different flows; (ii) the packets queued belong to the same flows and don't get therefore reordered within their flows; (iii) the additional introduced delay per packet is negligible, being of few clock cycles (in our prototype a packet would need to wait at most 6 clock cycles).

The hash tables used to store the flow contexts are implemented using RAM blocks. The EFSM tables are instead implemented using very small TCAMs. That is, an EFSM TCAM has 32 entries of 160 bits. Indeed, TCAM implementation over FPGAs is very inefficient and is currently a widely open research issue \cite{TCAM1,TCAM2,6665177}. Still, in Sec.~\ref{sec:usecases}, we will demonstrate that this small number of TCAM entries can easily capture the implementation of an iptables interface. The prototype fixes the parameters of the machine model as shown in Table~\ref{t:param}.

\begin{table}[t]
\small
\centering
\begin{tabular}{|c|c|c|}
 \hline
param. & value & descr.\\
\hline
k  & 4 & 
Number of flow context's registers. \\
& & Each register is 32bit long. \\
 \hline
& & 
Each condition is in the form \\
m & 8 & $var_1\ op\ var_2$, with operand being one \\
& & of $>$, $<$, $=$, and variables being \\
& & packet header's fields, registers or constants. \\
\hline
n & 32b & 
Size of packet's metadata moved between stages \\
\hline
h & 8 & 
Number or global registers. Each \\ 
& & register is 32bit long. \\
\hline
& & 
Number ALUs. Each ALU performs an \\
& & operation in the form $res = var_1\ op\ var_2$. \\
ALUs & 5 & $res$ and $var$s can be: global register, \\ & & flow context's register or packet's  \\ 
& & fields (including metadata). $op$ can be  \\ & & one of $+$, $-$, $shift$, etc. \\
\hline
\end{tabular}
\caption{Parameters of a OPP hardware prototype}
\label{t:param}
\end{table}

\begin{table}[t]
\small
\centering
\begin{tabular}{|c|c|c|c|}
\hline
resource type & Reference switch & OpenState & OPP switch   \\
\hline
\# Slice LUTs & 49436 (11\%) & 62637 (14\%) & 71712 (16\%)  \\
\hline
\# Block RAMs & 194  (13\%)  & 245 (16\%) & 393 (26\%) \\
\hline
\end{tabular}
\caption{Hardware cost comparison of OPP, NetFPGA SUME ref. switch and  OpenState.}
\label{t:synth}
\end{table}

The whole system has been synthesized using the standard Xilinx design flow. Table \ref{t:synth} reports the logic and memory resources (in terms of absolute numbers and fraction of available FPGA resources) used by the OPP FPGA implementation, and compare these results with those required for the NetFPGA SUME single-stage reference switch and a OpenState stage. We remark OPP is an evolution of the original OpenState stage. The synthesis results confirm the trend already shown by \cite{rmt}: the hardware area is dominated by memory, while adding intelligence/features in the logic require a small silicon overhead. 
Notice that the reported resources include the overhead of several blocks, such as the microcontroller for OPP configuration, the input/output FIFO for the 10GbE interfaces etc., which are required to operate the FPGA and do not need to be replicated for each stage. In fact, given the required resources, a NetFPGA SUME can currently host up to 6 stateful OPP Stages.

\vspace{-0.1in}
\section{The iptables use case}
\label{sec:usecases}
In this section, we present a possible implementation of the iptables interface using the OPP abstraction. While we support almost entirely the interface, for space constraints we limit the discussion to a relevant subset of functions. Nonetheless, we present the full implementation of three relevant use cases: a stateful firewall, a load balancer and a dynamic NAT. We begin the section with a short primer of iptables and then describe the three use cases.

\subsection{iptables primer}
Iptables is a well known Linux's user interface to control the \texttt{Netfilter} module, which is responsible for processing packets traversing the Linux's networking subsystem. In cooperation with the \texttt{conntrack} module, Netfilter supports a wide range of network functions such as: filtering, NAT, stateful firewall, load balancer, anomaly detection, etc.

Iptables uses rules to configure Netfilter. The first of the matching rules triggers the execution of a corresponding action. Rules are specified as follows:
\begin{lstlisting}
iptables $command $table $match $target 
\end{lstlisting}
\vspace{-0.05in}

\noindent where \textit{command} is, e.g., insert, delete, etc. and  includes also the specification of a Netfilter's \textit{hook}. A hook indicates the position within the IP networking subsystem, e.g., PREROUTING, FORWARD, POSTROUTING. Each \textit{table}, e.g., nat, filter, etc., provides a given set of packet processing capabilities. The \textit{match} is somewhat similar to an OpenFlow rule's match part, although being more flexible. Finally, the \textit{target} specifies the action to be applied, including: DROP, SNAT (source NAT), DNAT (destination NAT) and others.

We support almost completely the iptables interface. Nonetheless, we highlight that some functions have not been implemented, such as those that act on the packet's payload.

\subsection{iptables implementation}
Our iptables implementation is conceptually simple. A thin software layer translates the iptables rules into a set of entries for the OPP stages. The software layer is a regular OpenFlow controller extended to support OPP. In fact, we modify RYU\footnote{https://osrg.github.io/ryu/} and its OpenFlow implementation~\cite{bebaD24}, adding new protocol messages to populate and inspect EFSM tables, flow context tables, etc. 

The software layer most interesting part is probably in the way the translation is actually performed. In fact, the rule-based nature of iptables does not require any complex, e.g., programming language's compilation-like process. Instead, the translation is very often an almost 1-to-1 mapping with the OPP entries, provided that some relevant functional blocks are identified in advance. For example, the first use case implements the function for the Linux's conntrack module using a set of 7 EFSM table's entries. Therefore, in the rest of the section we focus on describing the translation from a relevant set of iptables rules to entries for the OPP stages. More specifically, using the RYU-powered iptables interface, we implement three different network functions combining 5 iptables' rules. In particular, we implement a stateful firewall (2 rules), a load balancer (2 rules) and a dynamic NAT (1 rule). Our implementation requires 4 OPP stateful stages, plus a stateless one, to support the three functions altogether\footnote{The number of stages may vary if a particular iptables function is used. E.g., the iptables' function RECENT actually defines a new table, hence, we need to allocate a new OPP stage to implement it.}. 

\begin{figure*}[!ht]
\centering
\includegraphics[width=1\textwidth]{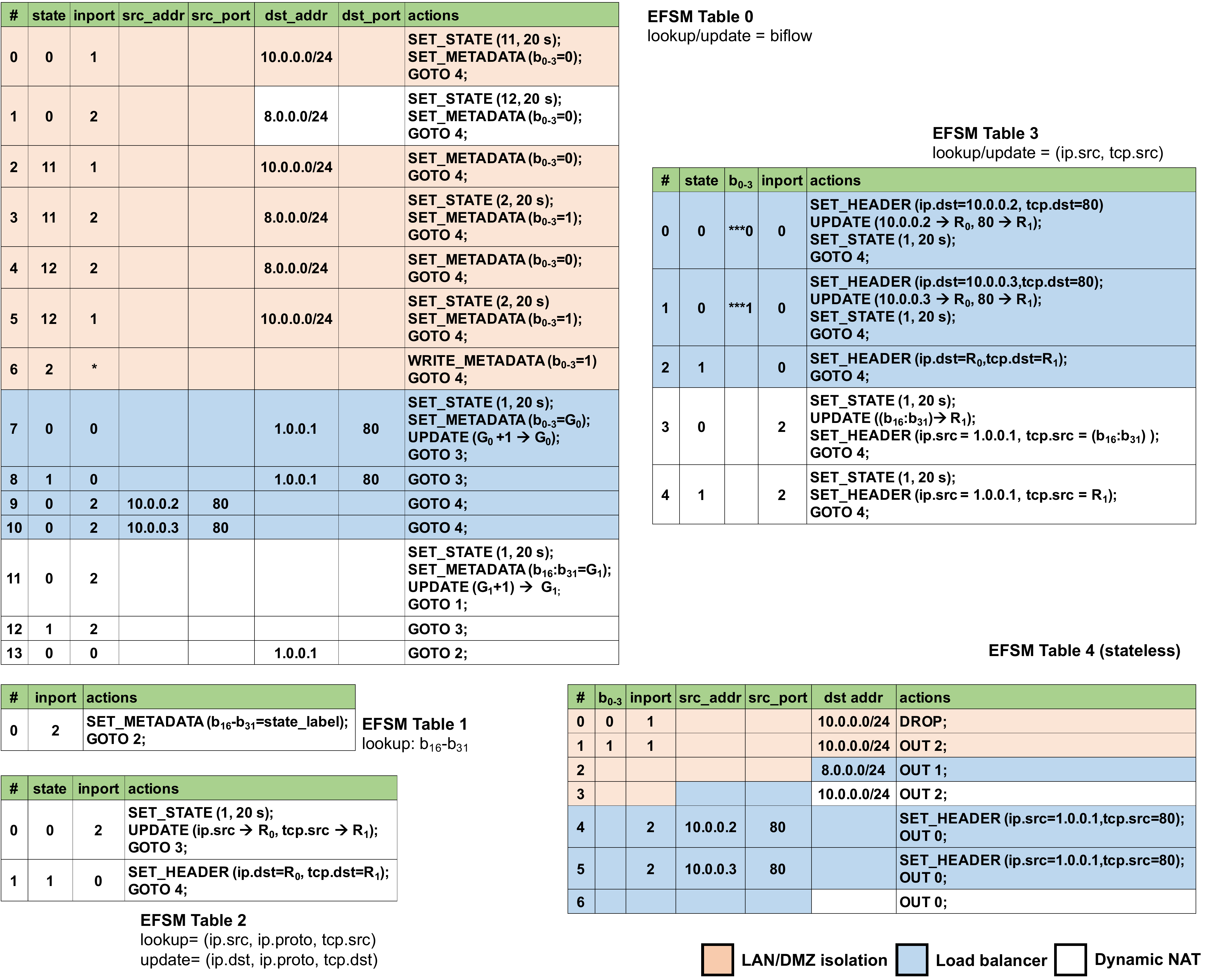}
\vspace{-0.1in}
\caption{OPP stages for the Iptables use cases}
\label{fig:opp_pl}
\vspace{-0.1in}
\end{figure*}

\begin{figure}[!ht]
\centering
\includegraphics[width=0.45\textwidth]{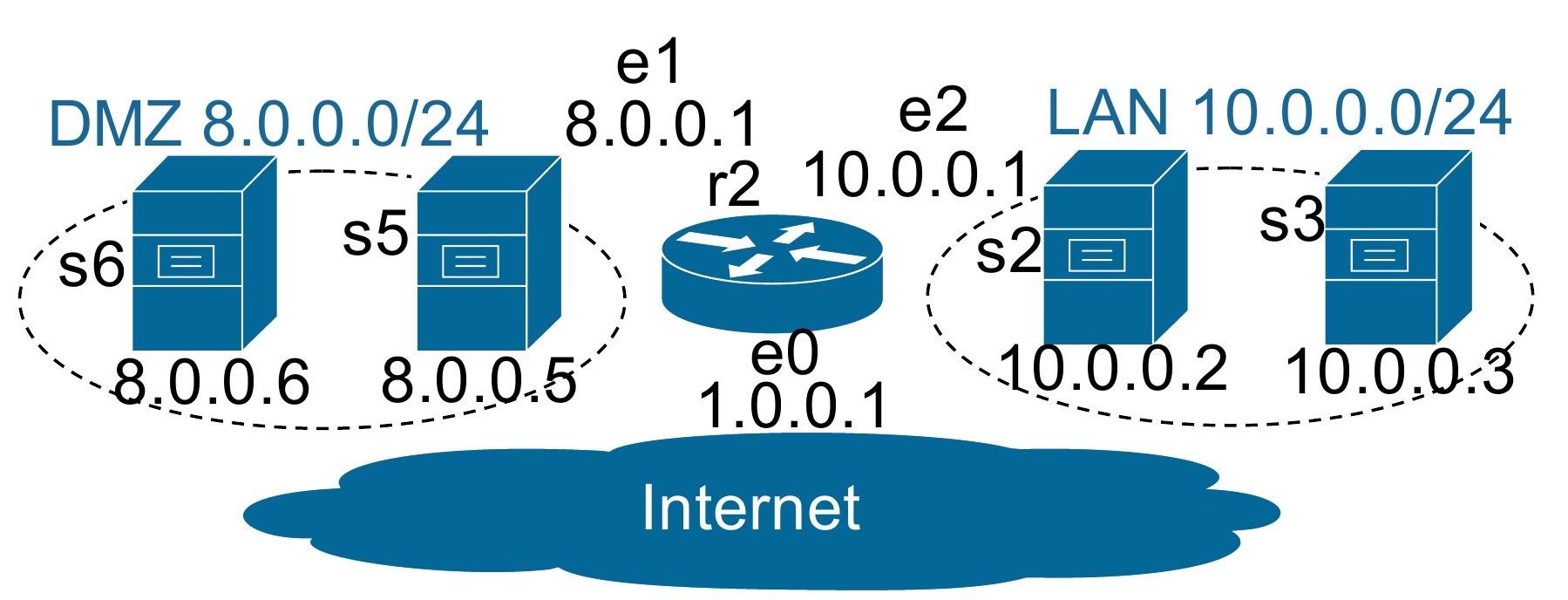}
\vspace{-0.1in}
\caption{Use cases scenario}
\label{fig:topo}
\vspace{-0.1in}
\end{figure}

To describe the OPP stages configuration, we represent just the EFSM tables and their entries, including the lookup and update key extractors configuration. Using the example scenario of Fig.~\ref{fig:topo}, the full configuration of our use cases is shown in Fig.~\ref{fig:opp_pl}. We assume that the node R of Fig.~\ref{fig:topo} is implemented as a VNF using OPP, with the network interfaces (e0, e1, e2) mapped on the OPP abstraction's ports (0, 1, 2). The colors identify the OPP entries used to implement the different use case. When an entry has multiple colors, it means that it is shared by the corresponding use cases. In the rest of the section, we describe the three use cases, one by one, in increasing level of complexity. 

\vspace{0.1in} \noindent  \textbf{LAN/DMZ isolation}. The first and most simple network function is a stateful firewall. The firewall allows a host in the DMZ to communicate with a host in the LAN only if the latter initiated the communication. 
In iptables, this is realised by the following rules:

\begin{lstlisting}
iptables -A FORWARD -i e2 -o e1 -j ACCEPT

iptables -A FORWARD -i e1 -o e2 -m state 
    --state ESTABLISHED -j ACCEPT
\end{lstlisting}
\vspace{-0.05in}

\noindent The translation of these rules only requires the use of stage 0 (7 rules) and of stage 4 (4 rules), which is stateless. Stage 0 is configured with a bidirectional 4 tuple flow\footnote{Source and destination addresses and ports.} for both the update and lookup key extractors. Such a configuration provides the same flow context's key regardless of the order of the address and port fields. I.e., both directions of the flow are mapped to the same flow context. The entries of stage 0 actually implement a connection tracking state machine, i.e., the Linux's conntrack function. When at least a packet per direction is exchanged, the connection is considered established. Such information is then passed to the stage 4 using the packet's metadata, where a forwarding decision is taken.

For example, assuming that the source \textit{10.0.0.2:123} (from the LAN) sends a packet to the destination \textit{8.0.0.5:678} (from the DMZ), the system would work as follows. The first packet from \textit{10.0.0.2:123} hits the stage 0, and not having any pre-existing flow context, is assigned with a default context identified by the lookup key \textit{\{10.0.0.2:123,8.0.0.5:678\}}. The default context's state label is set to 0, therefore, the EFSM table 0's entry \#1 is matched. The corresponding actions set a new state for the flow context (with state label value \textit{12}) and send the packet to stage 4. Here, notice that the flow context being updated is again \textit{\{10.0.0.2:123,8.0.0.5:678\}}, since the lookup and update keys are configured to be the same. In stage 4, the packet is forwarded towards its destination, to port 1 (EFSM table 4's entry \#2). 

The DMZ host's response packet will be associated with the same flow context of the first packet, having the same key \textit{\{10.0.0.2:123,8.0.0.5:678\}}. Here, recall we configured the key extractors to use a bidirectional flow hashing. 
The packet is associated with the already existing flow context, whose state label is set to 12. Therefore, EFSM table 0's entry \#5 is matched. The entry actions (i) update the state label to 2, which actually means "established", (ii) set the packet's metadata ($b_{0-3}$) to 1 and (iii) send the packet to stage 4. Here, the metadata value is used to signal that the flow is established, so the DMZ host's packet matches entry \#1 and is forwarded to the LAN. All the remaining packets belonging to this (bidirectional) flow will be associated with a state label 2 in stage 0, and then finally forwarded by stage 4's entries, effectively enforcing the firewall policy.

Notice that the flow context, set in stage 0's entries, is always associated with a 20 seconds idle timeout\footnote{The timeout values used throughout this section are just an example and are not motivated by any rationale.} (e.g., action $SET\_STATE(2, 20s)$). That is, when none of the processed packets is associated with such a flow context for more than 20 seconds, the context gets evicted.

\vspace{0.1in}
\noindent \textbf{Load balancer}. In this use case, we configure a load balancer function that assigns TCP connections to two web servers, in a round-robin fashion. Directing traffic to a given web server is as easy as configuring a static NAT rule. Nonetheless, the complexity of the use case is in the enforcement of the round-robin selection, and in the consistent forwarding after a connection has been assigned to a server. That is, the destination web server is selected when the first connection's packet is received, and all the remaining packets for that flow should be forwarded to that same web server.

In iptables, the following rules configure such function:

\begin{lstlisting}  
iptables -t nat -A PREROUTING -i e0
    -d 1.0.0.1 -p tcp --dport 80
    -m statistic --mode nth
    --every 2 -j DNAT
    --to-destination 10.0.0.3:80
    
iptables -t nat -A PREROUTING -i e0
    -d 1.0.0.1 -p tcp --dport 80
    -m statistic --mode nth 
    --every 1 -j DNAT
    --to-destination 10.0.0.2:80
\end{lstlisting}
\vspace{-0.05in}

\noindent Our implementation requires 2 stateful OPP stages (0 and 3) and a stateless stage (4) to translate the above rules. In stage 0, a global variable ($G_0$) is used to keep a counter that is incremented whenever a new flow starts. The value of the counter is then passed in the packet's metadata to stage 3. Here, the server assignment decision is taken looking at the last bit of the packet's metadata, which contains the counter value\footnote{Notice that the OPP hardware implementation does not support arbitrary modulo operations. Using the ESFM table (TCAM) mask feature, we implement a mod 2 operation. However, if we had to load balance among 3 servers, we could not use the same entries configuration shown in Fig.\ref{fig:opp_pl}. Fortunately, it is possible to emulate a Montgomery modular multiplication~\cite{montgomery1985modular} setting the right values for $G_0$ increments, together with a proper mask for the metadata matching in stage 3. Such a technique allows a programmer to describe, e.g., any arbitrary number of server in this use case.}. Depending on such value, a NAT operation is applied to the incoming flows from the public Internet, translating the destination address from the external facing address to one of the internal server's addresses. Finally, stage 4 performs the forwarding decision and applies the static NAT rules for the traffic in the opposite direction. I.e., it rewrites the source address, replacing the internal servers' addresses with the external facing address.

Considers now a source on the Internet, e.g., \textit{2.0.0.7:678}, sending its first TCP packet, to establish a connection to the load balancer's address \textit{1.0.0.1:80}. The packet will get to stage 0, where a default flow context will be assigned to it (with key \textit{\{2.0.0.7:678,1.0.0.1:80\}}). Therefore, it will match entry \#7, which applies four actions. First, the flow context is updated with a flow label $1$, which stands for "first packet received". Second, the value contained in $G_0$ is copied in the metadata ($b_{0-3}$). Third, the $G_0$ value is incremented ($UPDATE (G_0 + 1 \rightarrow G_0$)). Finally, the packet is sent to stage 3. Here, it is important to point that the second and third operation happen at the same time. In fact, the $G_0$ value before the increment is copied in the metadata (parallel read). 
In stage 3, the packet is associated with a new default flow context, identified by the key \textit{\{2.0.0.7:678\}}\footnote{Recall that each flow context exists only in the scope of the stage within which it was created. That is, each stage has its own flow context table and key extractors configuration.}. Assuming that the $G_0$ value copied in the packet's metadata was originally 0, the entry \#0 will be matched. The entry's actions (i) rewrite the destination address to one of the actual web servers (\textit{10.0.0.2:80}), (ii) copy corresponding address and port number in flow context's registers $R_0$ and $R_1$, (iii) set the flow context's state label to $1$, meaning "server assigned", and (iv) send the packet to stage 4. Here, the packet, with the rewritten destination address, matches entry \#3 and gets forwarded to the selected web server.

The server's response packet will be associated with a default flow context in stage 0. In fact, the packet is associated with a context key \textit{\{10.0.0.2:80, 2.0.0.7:678\}}. Notice that the first packet created instead a context with key \textit{\{2.0.0.7:678,1.0.0.1:80\}}. Therefore, entry \#9 is matched and the packet is sent to stage 4. We remark that since the entry's actions do not perform any flow context update, no entry for this context is created in the flow context table. In Stage 4, the packet matches entry \#4, that restores the source address to 1.0.0.1 before forwarding it to the Internet.

When a new packet from \textit{2.0.0.7:678} is received, it will be associated with the flow context \textit{\{2.0.0.7:678,1.0.0.1:80\}}, whose state label is $1$. The packet will match EFSM table 0's entry \#8, which sends it to stage 3. Here, the flow context with key \textit{\{2.0.0.7:678\}} is retrieved. This context's state label is $1$, as it was created by the first packet. Furthermore, flow context's registers $R_0$ and $R_1$ contain the assigned server address for such flow. Therefore, the packet matches entry \#2, whose actions rewrite the destination address and port with the values contained in the registers and send the packet to stage 4, where it is finally forwarded to the correct server. 

\vspace{0.1in} 
\noindent\textbf{Dynamic NAT}. 
The last function we present performs dynamic NAT between the LAN and the Internet of Fig.~\ref{fig:topo}, translating local source addresses into a public IP address, with a dynamically selected source port, and viceversa. The port is selected from a pre-configured bucket of available ports. In iptables, this is described with the following rule:

\begin{lstlisting}
iptables -t nat -A POSTROUTING 
    -i e2 -o e0 -j MASQUERADE
\end{lstlisting}
\vspace{-0.05in}
\noindent In this case, the rule translation uses 4 stateful stages plus the stateless stage. For the implementation of this function, however, we had to perform a special trick to represent the bucket of available ports. We use a dedicated OPP stage ($1$) and configure it to work as a memory stack. While the OPP machine model allows a programmer to play such a trick, we recognize that this is a stretch of the abstraction, which we will further discuss in Sec.~\ref{sec:discussion}. 

In particular, stage 1 is configured by the controller, using OpenFlow-like messages, to pre-populate the entries in the flow context table. The entries' state labels are in fact port numbers. The stage's lookup key extractor is configured to use a value contained in the packet's metadata. Such value is used as a stack pointer and should be set in a previous stage. Whenever a packet is received, the flow context pointed by the metadata value is attached to the packet. Such context's state label is in fact a port number. The port number is assigned to the packet's metadata by the matching entry's action, and therefore it can be used in a following stage. The controller runs a helper function periodically, monitoring the state of the stages, to track ports usage. The ports that became free (for instance because flow contexts were evicted) are pushed again the stage 1's flow context table.

Having clear the role of stage 1, we can now describe the other stages. Stage 0 is used to extract the "stack pointer" used in stage 1, to dynamically assign a port number. The pointer is stored in the global register $G_1$. After traversing stage 1, the packet's metadata will contain the assigned port number. Stage 2 is the only example, in the presented cases, in which a cross-flow state update is required. In effect, stage 2 is responsible for restoring the original NATted host's address and port number for packets coming from the Internet (port 0). This is done by storing the original values in two flow context's registers, when the first packet of a flow (coming from port 2) is received. However, the outgoing flow (from port 2) has to set such values for the incoming flow (from port 1). Therefore, stage 2's lookup key extractor is configured to use the source address and port of the packet, while the update key extractor uses the destination address and port. 
Stage 3 is instead in charge of performing the NAT for the outgoing flow. That is, it writes in the packet header the external facing address and the assigned port number. Finally, stage 4 performs the usual forwarding decision.

As in the previous case, we assume that the source \textit{10.0.0.4:123} is sending a packet to the destination \textit{2.0.0.1:678}. In stage 0, the packet is associated with a new flow context and matches entry \#11. The entry's actions are similar to the load balancer case, but this time a different global variable, $G_1$ is copied in the packet's metadata. In stage 1, the packet's metadata are used to lookup a pre-populated flow context. Such context's state label is copied in the packet metadata by the (only) matching entry and the packet is sent to stage 2. Here, a new context with key \textit{\{10.0.0.4:123\}} is associated to the packet, which therefore matches entry \#0. The entry's action does just an update of a flow context before sending the packet to stage 3. However, the flow context update happens on a different flow than the one to which the packet belongs. In fact, the update key extractor is configured to use the packet's destination address and port. Hence, while the packet is attached with the flow context \textit{\{10.0.0.4:123\}}, the action will update the flow context \textit{\{2.0.0.1:678\}}. The update sets the context's state label to $1$, meaning "translation entry stored", and sets the values of the origin host's address and port number in the $R_0$ and $R_1$ registers. The processing in stage 3 and 4 proceeds as in the previous cases, therefore we skip the description for brevity. Now, assume that the extracted port for the first packet was \textit{444}, then, a response packet from \textit{2.0.0.1:678} is sent to \textit{1.0.0.1:444}. The packet is associated with a new flow context in stage 0, and therefore it matches entry \#13, whose action sends the packet to stage 2. Here, the packet will be associated with the flow context \textit{\{2.0.0.1:678\}}, that was updated by the first packet. Therefore, the entry's actions (i) write the values stored in registers $R_0$ and $R_1$ in the packet's header destination address and port number, and (ii) send the packet to stage 4, where it is correctly forwarded to the LAN's host.

\vspace{-0.1in}
\section{Software implementation}
\label{sec:implementation}

While the previous section demonstrated the flexibility of the OPP abstraction, in this section we try to assess its scalability when implemented in software. We do so presenting a OPP software implementation that can scale to run on multiple CPU's cores. The section describes the implementation and the provided optimizations, including the support for multi-core processing. Finally, we report the results of experimental performance tests.

Our software implementation extends OFSoftSwitch~\cite{ofsoftswitch} (OFSS), an OpenFlow compliant software switch. OFSS is a popular tool in the academic community, as it provides a clean and flexible implementation of OpenFlow that makes it suitable for functional experimentation. However, the software was not originally designed with performance in mind. Therefore, in addition to implement the functional extensions required to support OPP, we optimized OFSS for performance.

In a nutshell, the architecture of OFSS consists of two processes, the one (\texttt{ofprotocol}) in charge of communicating with an external controller and to set general configurations, and the other (\texttt{ofdatapath}) implementing the actual switching operations. The \texttt{ofdatapath} module is designed as a single--process application and relies on the \texttt{netdev} library to access network devices. 

The reminder of this section elaborates upon the main performance impairments included in OFSS together with the countermeasures undertaken to remove or possibly mitigate their effect and accelerate the switch performance.

\subsection{Code Optimizations}
The first direction to improve the performance of OFSS was to virtualize its \texttt{netdev} library in order to integrate the I/O engine PFQ~\cite{jsac}, using its accelerated \texttt{pcap} adaptation layer. This allowed us to replace the slow \texttt{AF\_PACKET} sockets with the accelerated PFQ sockets without changing the OFSS code. Nonetheless, a thorough code analysis of OFSS revealed a significant number of possible performance bottlenecks. In the following, we list the major performance modifications to the original software architecture. 

\noindent\textbf{Dynamic Memory Allocation}. The datapath of OFSS makes extensive use of dynamic memory allocations and related memory releases. This dramatically impacts the packet forwarding performance as the cost of each pair of calls is around 200-500 CPU cycles. We implemented a \textit{zero--malloc} optimization that allows OFSS to generally run without performing dynamic memory allocations. 
Whenever required, the semantic of the data structures have been changed in order to cope with memory buffers without ownership, which are passed along the datapath as managed memory. Furthermore, the packet handler has been re-designed in order to fit into a single chunk of memory, replacing the original scattered model. This permits to save two extra additional memory allocations and deallocations. 

\noindent\textbf{Hash Maps Refactory}. Hash maps are pervasively used throughout the OFSS datapath. Wherever possible, hash maps have been replaced with more efficient \texttt{struct} data types to save very frequent memory indirections when accessing specific protocol fields. In addition, the remaining hash maps have been equipped with a set of managed small memory nodes, which are allocated at construction time. Size and number of such nodes are configurable at compile time, instead. Since hash tables are aware of whether the memory nodes are managed or not (using an annotation on each single node), they can concurrently use both the small set of pre--allocated nodes and the additional nodes allocated on--demand. In the case of the hash tables associated with the packet handler, this optimization allows to save (on average) up to 3 or 4 table rehashes, as they systematically host around a dozen of entries per each received packet.

\noindent\textbf{Zero Copy}. Both the semantics of \texttt{pcap} and PFQ allow one to take advantage of the memory persistency of a packet, during the call of a \texttt{pcap} handler. This semantic has been leveraged to retain from saving a copy of the payload of each packet, when not strictly required\footnote{ 
E.g., when the packet is consumed in the contest of the forwarding thread of execution.}. 
Strictly speaking, the \textit{zero--copy} optimization consists in removing a pair of \texttt{malloc/free} together with a \texttt{memcpy}.

\noindent\textbf{Batch processing}. The original version of OFSS processes one packet at a time. Instead, we enabled batch processing of packets. Therefore, the forwarding function of OFSS has been changed in order to consume, per each port, a batch of packets up to a configurable number, before switching to another port. The beneficial effects of such an optimization are mainly due to the increased cache locality that occurs while processing packets. In addition, in modern CPU, this mode of operation allows one to take advantage of packet pre--fetching.  That is the CPU is explicitly instructed to pre--fetch data while doing some other processing. As a result, the CPU can retrieve one or  more consecutive packets while the current one is being processed.

\subsection{Scalability}
One of the advantages of using PFQ is to enable fine--grained parallel computation in a simple and programmable way. Unlike other accelerating alternatives, PFQ integrates an in--kernel processing stage that is fully programmable through a high--level functional language. Such a processing engine is mainly intended as a ``pre--processing stage'' and allows the execution of \textit{dynamic} and \textit{hot--swappable} (i.e., atomically upgradable at run--time) processing pipelines (computations). Steering functions are particularly relevant to enable parallelism as they allow to deliver packets to group of sockets by using a hash based load balancing algorithm. More generally, steering can be performed according to arbitrary criteria with the overall target of distributing the processing and avoid state sharing across cores. Such feature turned out to be straightforwardly applicable to enable the scaling of our OPP software switch. In fact, we retrieve the PFQ's steering criteria (keys) directly from the \textit{lookup} and \textit{update} key extractor configurations.

\subsection{Evaluation}
We carried out an extensive experimental campaign, under different scenarios, to understand the absolute performance of our implemented prototype and its scalability. The experimental test bed consists of two identical PCs with 8-core Intel Xeon E5–1660V3 CPUs (3.0GHz), equipped with a pair of identical Intel 82599 10G NICs. One of the PCs runs the software switch the other is a load generator. Both systems run a Linux Debian stable distribution (kernel v. 3.16). A third server runs the controller and is connected to the switch's server using 1G control network interface. 

\begin{figure}[!ht]
\includegraphics[width=\columnwidth]{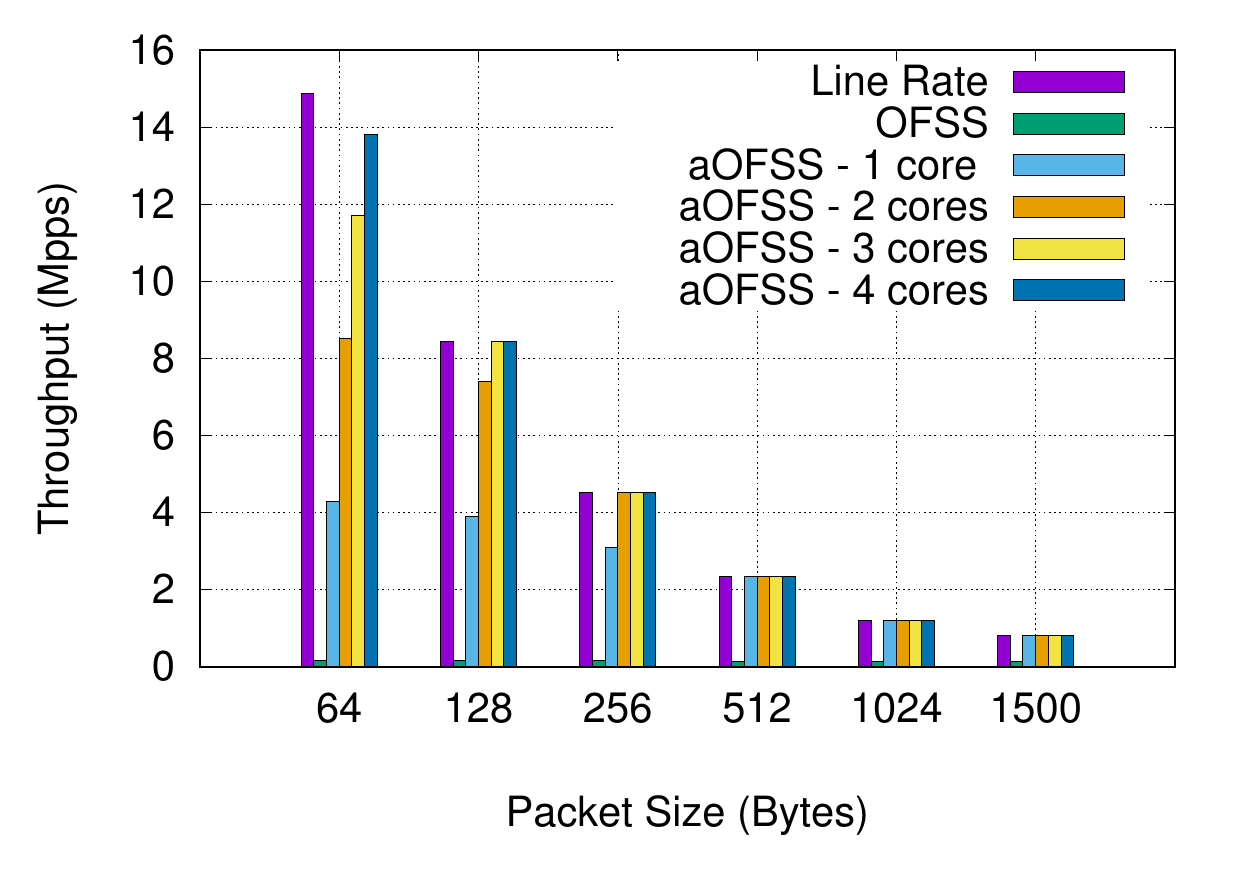}
\vspace{-0.4in}
\caption{OpenFlow pipeline throughput }
\label{speedtest}
\vspace{-0.1in}
\end{figure}

\begin{figure}[!ht]
\includegraphics[width=\columnwidth]{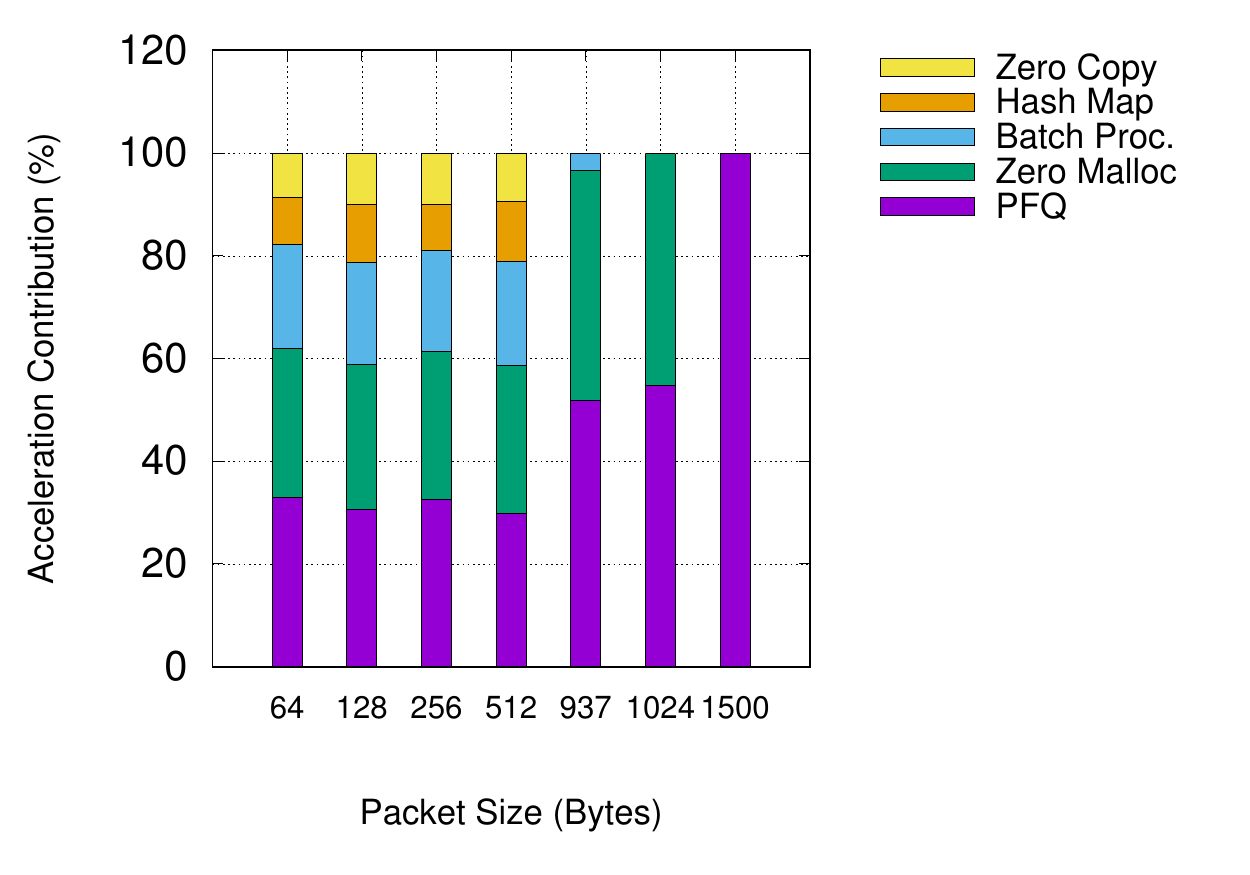}
\vspace{-0.4in}
\caption{Acceleration contributions to OFSS}
\label{accel-contrib}
\vspace{-0.1in}
\end{figure}

\noindent\textbf{OpenFlow performance}. The first set of experiments are pure \textit{speed tests} to benchmark the performance of the accelerated version of OFSS (aOFSS) when running a standard OpenFlow pipeline. Figure \ref{speedtest} shows the achieved throughput, when varying the number of cores and the packet sizes. Both the original OFSS performance and line rate limits are also reported for comparison. The implemented acceleration techniques provide a dramatic performance improvement, with the throughput nearly hitting line rate in all the cases. A up 90x speedup factor with respect to the original OFSS. 

The contribution of each optimization technique to the total throughput is reported in Figure \ref{accel-contrib}. The results are obtained by selectively switching off one contribution at a time on the accelerated version of OFSS running on a single core and measuring the observed performance drop. Data are finally normalized to give a fair visualization of each term as a stacked histogram. It is worth noticing that, in this case, multi--core acceleration is not accounted since the experiment was run on one core. The results show that the generic I/O acceleration provided by PFQ and the zero--malloc optimization have a mostly equally beneficial impact on the performance boost. For shorter packet sizes, the impact of the other optimizations is all but negligible as they contributes for up to 40\% of the overall performance improvement. 

\begin{figure}[!ht]
\includegraphics[width=\columnwidth]{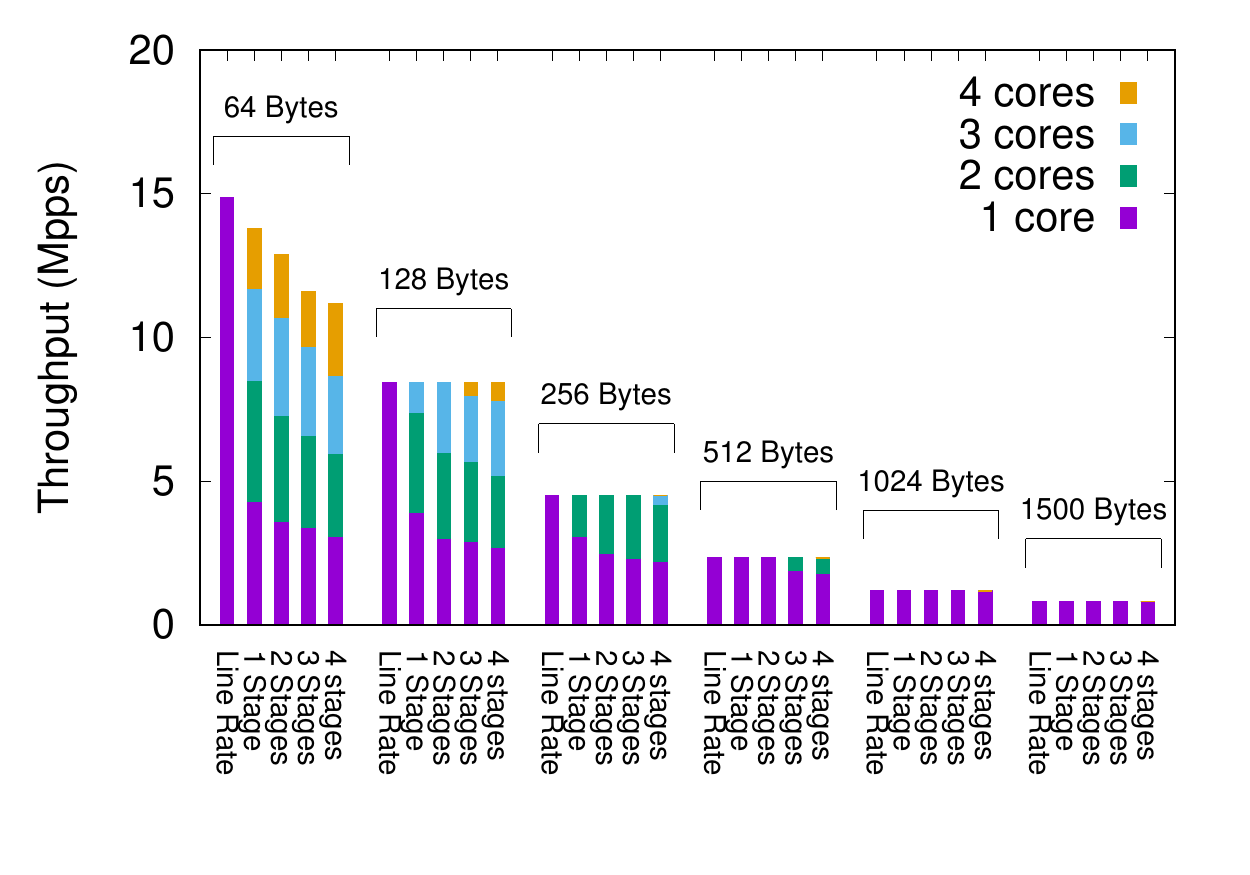}
\vspace{-0.5in}
\caption{Stateless OPP stages throughput}
\label{stateless-stages}
\vspace{-0.1in}
\end{figure}

\noindent\textbf{OPP performance}. The second set of experiments measures the aOFSS performance when using OPP, including multiple pipeline configurations. Figure \ref{stateless-stages} shows the throughput achieved by aOFSS when using \textit{stateless}  OPP stages. The system still hits line rate for packets of realistic sizes bigger than 128B. In the most critical case of shortest packet size, performance decreases linearly with the number of stages, but still reaching well above 10 Mpps with 4 running cores. Notice that the performance for one stage are comparable to the ones of OpenFlow. In fact, a stateless OPP stage is functionally equivalent to an OpenFlow table.

\begin{figure}[!ht]
\includegraphics[width=\columnwidth]{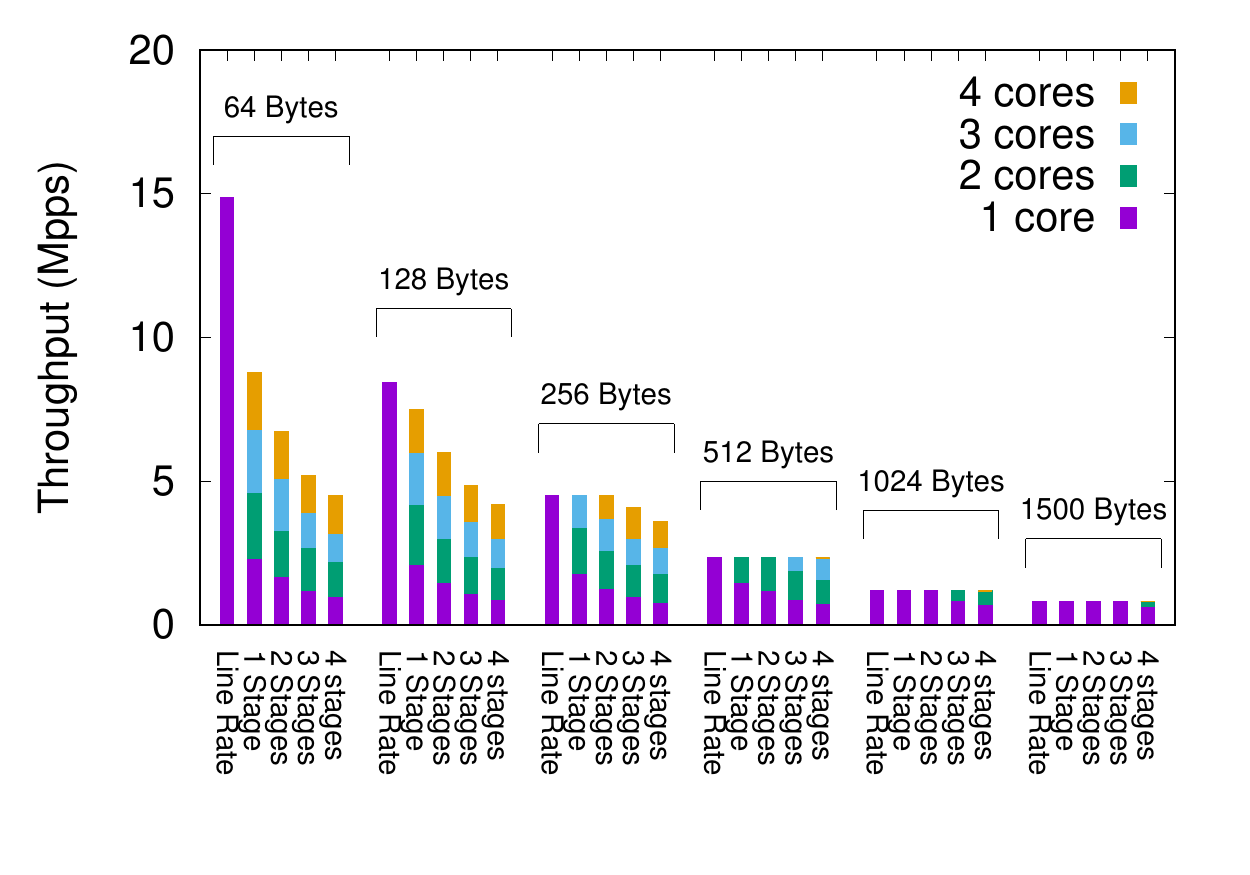}
\vspace{-0.5in}
\caption{Stateful OPP stages throughput}
\label{stateful-stages}
\vspace{-0.1in}
\end{figure}

Figure \ref{stateful-stages} shows the performance for a pipeline of \textit{stateful} OPP stages. As expected, the performance decreases, with line rate achieved for packet sizes bigger than 256B. The degradation is more significant as the number of stages increases. However, we remark that in our test we measured a somewhat worst--case behavior. That is, \textit{every packet performs a state update}. 
As shown in Sec.~\ref{sec:usecases}, in real use cases most of the packets perform just a state lookup, with only few of them actually triggering a state modification. 

Finally, we note that the above results provide a lower bound for the previously described use cases, namely LAN/ DMZ isolation (1 stateful stages + 1 stateless stage), load balancer (2 stateful stages + 1 stateless stage) and dynamic NAT (4 stateful stages + 1 stateless stage). However, we did not test the actual use cases' rules, furthermore, the tests did not include the evaluation of operations on global variables. Testing the use cases performance, including the operations on global variables, would have required the testing of actual traffic traces to be meaningful. Since we lacked significant traffic traces for the presented use cases, we leave the evaluation of the impact of such operations for future work.

\vspace{-0.1in}
\section{Discussion}
\label{sec:discussion}
In this section we discuss the advantages and the limits of the OPP abstraction when using it for realizing scalable (virtual) software network functions, report on lessons learned and possible future work.

\vspace{0.1in}
\noindent\textbf{Abstraction flexibility}. 
Apart from the presented use cases, iptables allows a network administrator to express fairly complex network functions. We implemented an almost complete iptables' interface using OPP and a number of network functions. Therefore, we are quite confident that the OPP abstraction can capture a large number of functions. Furthermore, although not presented in the examples, we could implement also functions that required packet generation instructions, since our prototype integrates the InSP API~\cite{insp}.

\noindent\textbf{Software scalability}. The OPP machine model can be efficiently implemented in software. While a hardware implementation that provides line rate performance is available for many abstractions, a scalable software implementation builds on completely different requirements. The concept of \textit{flow state} greatly helps the scaling of the OPP machine model in software, allowing the distribution of the load to different cores. A feature we could straightforwardly implement in the majority of the cases using standard software acceleration libraries. 

However, it is worth noticing that the access to the global variables is likely to introduce an important performance overhead in software implementations. Unfortunately, we admittedly did not test the performance impact in presence of global variables operations. 
As a general rule, a programmer should always try to express her algorithms avoiding the use of global variables. Otherwise, she should be well aware of these overheads. For example, in the load balancer example of Sec.~\ref{sec:usecases}, a global variable is used only on the first packet of a new flow. Hence most of the traffic is not introducing the mentioned overheads. Indeed, a traffic pattern with a large number of new starting flows in a short time, as it may be the case during a Denial of Service (DoS) attack, could severely impact the overall system forwarding performance. Notheless, we point that such an issue is not specific to our implementation, and is typical for this kind of systems~\cite{lbdos}. In fact, a form of DoS detection and mitigation function is usually deployed to protect the system by such attacks. Incidentally, our iptables implementation can deploy such a protecting network function, on the very same pipeline, using the proper set of iptables rules.

\noindent\textbf{Software performance}. Our software prototype does not shine in terms of absolute performance numbers, when compared to the state of the art. However, our work did not aim at improving the state of the art in software acceleration. In fact, we selected a not optimized software switch for our prototype. Our focus was to study weather the selected abstraction, OPP, was suitable for scaling in multi-core systems, and why. Furthermore, we notice that our system cannot be fairly compared to traditional software switches, which apply mostly "stateless" processing. Systems such as SoftFlow~\cite{softflow} are a more fair comparison term.

\noindent\textbf{Slow path}. In addition to the rules translation, our iptables implementation requires the controller to perform also some helper tasks on the slow path. In a production environment, we expect such tasks to be run by the different VNFs that use OPP as a mean to implement their fast paths.

\noindent\textbf{Programming complexity}. Programming functions in OPP is hard. Or at least, it is not straightforward as it could be with a high-level language. We believe that approaches such as Domino are very promising to address the issue. Introducing support for specifying flow-level consistency requirements in such languages is indeed an interesting area for future research.

\noindent\textbf{Abstraction maturity}. While some things are done right in OPP, others have still space for improvements. We provide two examples. First,  ALUs support the same operations on global registers and on flow context's registers. In hardware implementations, the complexity of operations on global registers is limited by the need to execute them atomically in one clock cycle. With the flow state concept, such requirement is relaxed for operations on flow context's registers. Supporting more complex operations at the flow level could open up the space for additional use cases. Second, we had to use a full stage in the dynamic NAT example to implement a memory stack. Such a function could be probably better provided as a dedicated function in the abstraction, e.g., implemented as an action. 
These two examples motivate our intention to further refine the OPP abstraction in future.

\noindent\textbf{FPGA implementation}. The most FPGA's resources consuming blocks of an OPP stage are those used for the TCAM (EFSM table) and the hash table (flow context table). In particular, the TCAM is generated only using the FPGA logic blocks and flip-flop. With the current FPGA technologies, we expect to be technically limited to 32-64 TCAM entries per OPP stage. Possible improvements could be achieved by using suitable FPGA tailored TCAM implementations such as the one recently presented in~\cite{6665177}. Inclusion of TCAM memories on chip for FPGA-based smart NICs is also a possible future option that would drastically tackle the issue. Instead, the scaling issues of hash tables is easier. Already today, the OPP prototype, based on the last generation NetFPGA board, could theoretically provide up to 13 OPP stages, with 16k entries of 256b for each stage. Furthermore, the new FPGAs, such as the Xilinx Virtex UltraScale+ family~\cite{virtexUltraScale}, can provide much larger memory blocks.

\noindent\textbf{Implementation options}. While we leveraged an FPGA-based hardware implementation and provided a software implementation, it would be interesting to verify how different architectures would implement the OPP abstraction. For instance, using NPUs, GPUs or even ASICs.

We also tried to described the OPP pipeline using P4. Unfortunately, as mentioned in Sec.~\ref{sec:abstractions}, at the time of writing the language could not capture the required consistency models. In fact, in \cite{pisces} the authors do not implement stateful operations when compiling P4 to a software switch target.

\vspace{-0.1in}
\section{Related Work}
We discussed stateful forwarding abstractions already in Sec.~\ref{sec:abstractions}. In this section, we briefly review related work on NFV, the implementation of high-performance functions, related abstractions and issues in supporting the virtualization and sharing of hardware accelerators.

\vspace{0.1in}\noindent 
\textbf{VNFs implementation}. SoftFlow~\cite{softflow} is probably the closest related work. In SoftFlow, OpenVSwitch is extended to integrate more flexible processing blocks called SoftFlow actions, which can implement complex stateful functions. Combining SoftFlow actions with an OpenFlow-like pipeline of MATs enables a developer to implement arbitrary network functions.  However, each SoftFlow action is in fact a black box, which consumes and produces packets, i.e., as if it was a VM attached to an OpenFlow switch. Conversely, in our case network functions are entirely programmed using a white box approach based on the EFSM abstraction. In fact, SoftFlow can only offload packet classification to hardware NICs and just for packets entering the system, while we can potentially offload both packet classification and stateful operations to the smart NIC.

Click~\cite{click} adopts a model in which arbitrary functional blocks, called elements, can be composed into graphs to implement a network function. ClickNP~\cite{clickNP} uses the Click's abstraction but adds the possibility to implement some elements as hardware functions to be run on, e.g., a smart NIC. However, in SoftFlow, Click and ClickNP actions or elements implementation is still a complex task, which has to be performed if there are no pre-implemented modules that meet the developer's need. 
To address this issue, NetBricks~\cite{NetBricks} defines as abstraction a set of more fine-granular primitives that combined can describe a large number of software network functions. The primitives implementation is optimized, therefore functions expressed using the NetBricks' abstraction provide high-performance. In this sense, our approach is similar, since we adopt the set of fine-granular MAT-based OPP functions to describe network functions. Still, the approaches differ in flexibility and hardware support. NetBricks is more flexible and expressive, but targets pure software functions. Our OPP-based solution can express only network functions that deal with packet headers, but provides full hardware support.

\vspace{0.1in}\noindent 
\textbf{Hardware acceleration}. 
Using hardware acceleration to meet performance requirements has been explored in the past for specific cases, e.g., firewalls implementation~\cite{npfirewall}. Recently, the increase in network speed, combined with the trend of NFV, is fueling a new stream of research to support the hardware acceleration features provided by the NICs. For example, Dragonet~\cite{Shinde:2013:WNT:2490483.2490484} is an OS network protocol stack that takes into account NIC's hardware capabilities for the implementation of network protocols. Our work goes in a similar direction, exploring a different part of the solution space: we select a less flexible abstraction that captures only a subset of the NIC's hardware capabilities. Nonetheless, such abstraction is easier to leverage for the programming of network functions and has readily available implementations. 

FlexNIC~\cite{flexNIC} envisions the support of RMT in future NICs, providing a way to execute the RMT-based processing while exchanging packets between the NIC and the host's memory. We consider such work orthogonal to our contribution, since the NIC could use an OPP-like processing model instead of one based on P4. 

\vspace{0.1in}\noindent 
\textbf{Software acceleration}.
An extensive comparison of software accelerated capturing techniques can be found in~\cite{braunimc2011,moreno2015,gallenmuller2015}. Relevant software accelerated engines are PF\_RING~\cite{pfring_imc}, PF\_RING ZC (Zero Copy)~\cite{pfringdna}, Netmap~\cite{netmap}, DPDK~\cite{dpdk} and PFQ~\cite{jsac}. PF\_RING ZC, Netmap and DPDK bypass the Operating System by memory mapping the ring descriptors of NICs at user space, allowing even a single CPU to receive 64 bytes long packets up to full 10 Gbps line speed. In addition, DPDK adds a set of libraries for fast packet processing on multicore architectures for Linux. 
Netmap and DPDK have been successfully used in accelerating soft switch as in the case of the VALE~\cite{vale} switch and mSwitch~\cite{mswitch} (netmap) and CuckooSwitch~\cite{cuckoo} and DPDK vSwitch~\cite{vswitch} (DPDK). Netmap was also used to accelerate packet forwarding in Click~\cite{rizzo-click}. PFQ, instead, relies on vanilla device drivers and leverages different levels of parallelism to accelerate packet I/O. In addition, PFQ is equipped with a native functional language to program in--kernel early stage packet processing~\cite{ANCS}.

\section{Conclusion}
In this paper, we demonstrated that an abstraction that defines flow-level states can be efficiently implemented in software. Furthermore, for functions that work only with flow-level states, such implementation can be easily scaled to run on multi-core systems.

\vspace{0.05in}
The OPP software implementation is available at \url{github.com/beba-eu/beba-switch}. We plan to upstream the performance enhancement to OFSoftSwitch. A subset of the OPP API (mainly OpenState) is currently under discussion for inclusion in OpenFlow v.1.6. 

\bibliographystyle{abbrv}
\bibliography{biblio}

\end{document}